# Superconducting $I\bar{4}3m$ CSH$_7$ model applied to resistive transition temperature data for compressed C-S-H at high pressure

Dale R. Harshman [1,a)] and Anthony T. Fiory [2]

[1] Physikon Research Corporation, Lynden, WA 98264, USA
[2] Bell Labs Retired, Summit, New Jersey 07901, USA

**ABSTRACT**

This article updates version 1 by restricting consideration to only the resistive data and excluding the questioned 287.7-K datum reported for carbonaceous sulfur hydride in Snider *et al.*, Nature **585**, 373 (2020). The superconducting transitions are considered in terms of the theoretically-discovered compressed $I\bar{4}3m$ CSH$_7$ structure of Sun *et al.*, Phys. Rev. B **101**, 174102 (2020), which comprises a sublattice similar to $Im\bar{3}m$ H$_3$S with CH$_4$ intercalates. Positing an electronic genesis of the superconductivity, a model is presented in analogy with earlier work on superconductivity in $Im\bar{3}m$ H$_3$S, in which pairing is induced via purely electronic Coulomb interactions across the mean distance $\zeta$ between the S and H$_4$ tetrahedra enclosing C. Theoretical superconducting transition temperatures for $I\bar{4}3m$ CSH$_7$ are derived as $T_{C0} = (2/3)^{1/2}\sigma^{1/2}\beta/a\zeta$, where $\beta = 1247.4$ Å$^2$K is a universal constant, $\sigma$ is the participating charge fraction, and $a$ is the lattice parameter. Analysis suggests persistent bulk superconductivity with a pressure-dependent $\sigma$, increasing from $\sigma = 3.5$, determined previously for $Im\bar{3}m$ H$_3$S, to $\sigma = 7.5$ at high pressure owing to additionally participating C-H bond electrons. With $a$ and $\zeta$ determined by theoretical structure, calculations of $T_{C0}$ at the highest pressures, 258 and 271 GPa, are in agreement with resistive transitions to within an overall uncertainty of ± 3.5 K.

---

## I. INTRODUCTION

Resistive transition temperatures ($T_C$) in Snider *et al.* [1] are indicative of near room temperature superconductivity in compressed carbonaceous sulfur hydride (C-S-H) [2]. This article updates a prior analysis [3] by restricting consideration to only the resistive data for $T_C$, excluding the susceptibility results that were subsequently found to be of questionable value [4-7], and notes that the maximum $T_C = 287.7 ± 1.2$ K is itself an individually questioned datum [8,9].* The $T_C$ of C-S-H exceeds the previous high-$T_C$ record of $Fm\bar{3}m$ LaH$_{10}$ [10-12] and approaches the ambient-temperature regime [13].

Pressure dependence in $T_C$ suggests a possible evolution in superconductive dominance between differing superconducting phases. Of particular note are the anomalously narrow transition widths $\Delta T_C$ in most of the resistance transitions, indicative of a continuum of superconducting states. A narrow $\Delta T_C$ also persists in high magnetic field, which may be evidence of strong pinning or trapping of magnetic fluxons [14,15]. Although alternative explanations were offered for the anomalous behavior observed [16-18], strong-coupled bulk superconductivity is assumed for the present work.

While the precise composition and structure of C-S-H are not determined in the original work, unity ratio of elemental C to S and Raman analysis of non-metallic precursor crystals have suggested CSH$_x$ stoichiometry and possibly H$_2$-containing compounds such as (CH$_4$)$_x$(H$_2$S)$_{2-x}$H$_2$ [1]. Analysis of single-crystal x-ray diffraction of a sample with similar preparation ($T_C$ unreported) has determined an orthorhombic system and ruled out $I4/mcm$ symmetry [19]. For C-S-H synthesized from sulfur and methane precursors and followed by laser annealing, x-ray and Raman analyses find formation of an orthorhombic $Pnma$ structure, which has fostered the suggestion that a stable sulfur-based cage structure is a host of superconductivity analogous to $Im\bar{3}m$ $H_3S$ [20].

---

*Typographical errors in Ref. 3 are also corrected.

Several theoretical treatments have considered thermodynamically stable forms of carbon-doped $H_3S$ [21-23]. However, accurate calculations of electron-phonon interactions predict values of $T_C$ that fall significantly below experiment, indicating a theoretical breakdown suggestive of a novel mechanism for the superconductivity in C-S-H [23]. Alternatively, there are several theoretical compressed phases of $CSH_7$, which include metastable structures of low formation enthalpy (~0 meV) [23-25]. Discovered computationally [24], $I\bar{4}3m$ $CSH_7$ comprises a host H-S-H chain framework similar to $Im\bar{3}m$ $H_3S$, but with enclosed intercalated $CH_4$ molecules [24], while orthorhombic $Pnma$ $CSH_7$ has 2D characteristics with layers in the $bc$ plane [25]. Electron-phonon calculations predict relatively low $T_C$ of 181 K at 100 GPa for $I\bar{4}3m$ $CSH_7$ [24], and $T_C$ of 157 − 170 K at 150 GPa for $Pnma$ $CSH_7$ [25], diminishing at higher pressure in each case. Host-guest hydride structures are also explored theoretically in a recent phonon-based study of high-pressure phases of $CS_2H_{10}$, in which both the $Cmc2_1$ and $P3m1$ phases have $CH_4$ molecules inserted into the $Im\bar{3}m$-$H_3S$ and $R3m$-$H_3S$ sublattices, respectively; results indicate a detrimental effect of intercalated $CH_4$ on super-conductivity, in contrast to experiment [26].

Following the stoichiometry suggested by experiment [1], superconductivity with highest $T_C$ is assumed to occur in a nominally $CSH_x$ material. Rationalized by structural study [20], as well as by findings of quantitative agreement with experiment, analysis is centered on the theoretical $I\bar{4}3m$ $CSH_7$ structure, comprising a cubic unit cell containing eight $(H_3S)(CH_4)$ formula units [24]. In lieu of electron-phonon coupling, which systematically disagrees with experiment, a purely electronic mechanism, in analogy with that described for $Im\bar{3}m$ $H_3S$ [27], is considered to be the origin of the extraordinarily high values of $T_C$, in which pairing is mediated via electronic interactions between charges of the H-S-H network and those of intercalated $CH_4$. Originally introduced for layered structures in Ref. 28 and subsequently extended to 3D systems including the superhydrides $Im\bar{3}m$ $H_3S$ and $Fm\bar{3}m$ $LaH_{10}$ [27,29], the model treats superconductive pairing as arising from electronic interactions between two charge reservoirs, denoted type I (superconducting) and type II (mediating), with a calculated transition temperature defined as $T_{C0} \propto (e^2/\zeta)(\sigma/A)^{1/2}$. Here, $\sigma$ is the charge fraction in each reservoir, being equal at optimal $T_{C0}$, $\zeta$ is the distance separating the reservoirs, and $A$ is the type I area [28]. Structural analogy between $I\bar{4}3m$ $CSH_7$ and $Im\bar{3}m$ $H_3S$ logically suggests that the superconducting condensate in C-S-H resides in the extended H-S-H host sublattice network (see Ref. 27), while the $CH_4$ molecules form the secondary sublattice, thus establishing the charge reservoirs as types I and II, respectively.

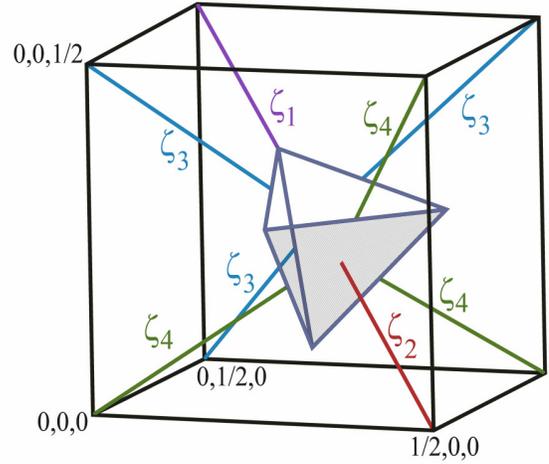

**FIG. 1.** Schematic structure of $I\bar{4}3m$ $CSH_7$, indicating the four distances $\zeta_1$, $\zeta_2$, $\zeta_3$, and $\zeta_4$ for calculating the average $\zeta$, the $H_4$-tetrahedron, and coordinates of S ions located at cube corners.

Determining the charge fraction $\sigma = 15/2 = 7.5$ from the maximum participating valence electrons of C, S, and H, divided equally between the two charge reservoirs, and $\zeta$ from the average distance between an S site in the H-S-H host sublattice and the nearest points on the eight neighboring $H_4$-tetrahedra of the $CH_4$ intercalants, the maximum transition temperature is calculated to be $T_{C0} = 283.0 \pm 0.7$ K for $I\bar{4}3m$ $CSH_7$ at 265 GPa; uncertainty corresponds to the theoretical extrapolation of the pressure dependence (Appendix). This result agrees remarkably well with experiment. The narrow $\Delta T_C$ occurring over a wide pressure range and the concomitant behavior of $T_C(P)$ [see Figs. 1(a) and 1(c) of Ref. 1, respectively] indicate not only a broad continuum in charge equilibrium, but also a strong pressure-dependent participating charge fraction $\sigma(P)$ varying in the range $3.5 \lesssim \sigma \lesssim 7.5$ attributable to the electrons of $CH_4$. Since $\sigma = 3.5$ for $Im\bar{3}m$ $H_3S$ [27], essentially all of the pressure variation in $\sigma(P)$ can be attributed to the $CH_4$ electrons.

In this work, notation $T_C$ refers to experimental data and $T_{C0}$ designates the results of the theoretical calculations. Section II presents the theoretical model for electronic mediated superconductivity in $I\bar{4}3m$ $CSH_7$, with determination of maximum $T_{C0}$, and analysis of $\sigma(P)$ and $T_{C0}(P)$ assuming pressure-dependent dispersion of the C-H bonds. In Sec. III important features of the experimental data of Ref. 1 are analyzed, seemingly conflicting x-ray data are addressed, and supporting evidence for non-phononic superconductivity in the superhydrides, along with a comparison with other optimally-compressed high-$T_C$ materials, are provided. Summary and concluding remarks are given in Sec. IV. An Appendix is included describing the methodology of extracting the pressure dependence of the structural parameters for $I\bar{4}3m$ CSH$_7$ and associated errors.

## II. THEORETICAL MODEL

The inter-reservoir electronic interaction model for high-$T_C$ superconductivity has enjoyed significant success in accurately predicting the transition temperatures of numerous materials optimized to yield maximal $T_C$, including most recently $Im\bar{3}m$ H$_3$S and $Fm\bar{3}m$ LaH$_{10}$ [27,29]. The calculated transition temperature is given generally by the algebraic relation,

$$T_{C0} = k_B^{-1} \Lambda [\sigma \eta / A]^{1/2} e^2 / \zeta, \quad (1)$$

where $\Lambda$ (= 0.007465 Å) is a universal constant, $\sigma$ is the participating charge fraction per formula unit (f.u.) from the valence electrons of constituent atoms (e.g., C, S, and H), $A$ is the surface area per f.u. of the (type I) charge reservoir hosting the superconducting condensate, $\eta$ (equal to unity for $I\bar{4}3m$ CSH$_7$) is the number of type II layers (see, e.g., Ref. 28 for the definition of type II layers and $\eta$), and $\zeta$ is the mean distance between the interacting species in the two reservoirs.

The theoretical compound $I\bar{4}3m$ CSH$_7$ of Ref. 23, comprising an H$_3$S-like host sublattice containing intercalated CH$_4$, is found through analysis presented herein to be demonstrably applicable to the compressed C-S-H studied in Ref. 1. Drawing on the structural correspondence with the dual-sublattice structure of $Im\bar{3}m$ H$_3$S, the H$_3$S-like framework is modeled as type I, making the CH$_4$ intercalates type II. Analogous to interaction distances determined for compounds comprising polyhedral structures, e.g., in $Fm\bar{3}m$ LaH$_{10}$ and Cs$_3$C$_{60}$ [29,30], $\zeta$ for $I\bar{4}3m$ CSH$_7$ is the mean nearest distance between an S (type I) site and the eight neighboring H$_4$-tetrahedra of the CH$_4$ (type II) intercalate. Area $A$ in Eq. (1) is formulated similarly as for $Im\bar{3}m$ H$_3$S by taking the surface area of the cube enclosing one f.u. [27], yielding $A = (3/2)a^2$ in terms of cubic lattice parameter $a$, given eight f.u. per unit cell.

Expressing theoretical $T_{C0}$ by defining the factor $\beta \equiv k_B^{-1} e^2 \Lambda$ (=1247.2 K–Å$^2$) provides a compact representation of Eq. (1), convenient for drawing comparisons to experiment. Thus, in terms of $\sigma$, $\zeta$, and $a$, Eq. (1) is evaluated for $I\bar{4}3m$ CSH$_7$ as,

$$T_{C0} = (2/3)^{1/2} \sigma^{1/2} \beta / a \zeta . \quad (2)$$

The charge fraction $\sigma$ in Eq. (2) is determined theoretically by the participating valence electrons of the constituent atoms in one f.u., a number that is divided by 2 for sharing between the type I and II reservoirs (a factor denoted $\gamma = 1/2$) [28]. The value $\sigma = 3.5$ was previously determined for $Im\bar{3}m$ H$_3$S from analysis of calculations for hydrogen- and sulfur-projected density of states, showing a total of 7 participating electrons from S $3p^4$ and $3 \times$ H $1s^1$ and non-participating electrons from core-like S $3s^2$ [27]. Since the S-projected density of states for $I\bar{4}3m$ CSH$_7$ quantitatively resembles that of $Im\bar{3}m$ H$_3$S, at least within several eV of the Fermi level [24], the participating valence electrons for $I\bar{4}3m$ CSH$_7$ is a maximum of 15, as counted from C $2s^2 2p^2$, S $3p^4$, and $7 \times$ H $1s^1$, predicting a charge fraction elevated to a maximum of $\sigma = 15/2 = 7.5$.

In view of experimental indications of optimization occurring over a wide pressure range, values of $\sigma$ in Eq. (2) are hence modeled with a fixed number from H$_3$S, augmented by a pressure-dependent number of participating CH$_4$ electrons. Since experiment points to equivalence between measured $T_C$ and theoretical $T_{C0}$ at various pressures, experimental determinations of charge fractions, denoted as $\sigma^{(X)}$, are derived from experiment according Eq. (2) by assuming theoretical determinations of $a$ and $\zeta$. Sections II. A and II. B describe further the calculations of $a$ and $\zeta$ from theoretical pressure dependence in lattice spacings for $I\bar{4}3m$ CSH$_7$ [24], the model introduced for the pressure dependence in $\sigma$, results for $\sigma^{(X)}$ derived from experimental $T_C$, and overall comparisons between theoretical $T_{C0}$ and experimental $T_C$ as functions of pressure.

### A. Calculation of maximum $T_{C0}$

According to the theoretical structure of $I\bar{4}3m$ CSH$_7$ at $P = 100$ GPa [24], the H$_4$-tetrahedra, comprising Wyckoff positions H(8c) and H(24g), have an apex or a face nearest to S(2a) sites and edges nearest to S(6b) sites. The cubic unit cell per f.u. is illustrated in Fig. 1, where cube corners represent S sites and an enclosed tetrahedron represents H$_4$. Nearest distances between a cube corner and sites on the tetrahedron surface are indicated as $\zeta_1$ and $\zeta_2$ for an apex and a face center, respectively, and $\zeta_3$ and $\zeta_4$ for the two three-fold symmetric edge centers. Using theoretical structural parameters [24], the calculated distances, in terms of lattice parameter $a$, are $\zeta_1 = 0.2950a$, $\zeta_2 = 0.3458a$, $\zeta_3 = 0.3343a$, and $\zeta_4 = 0.2901a$. Averaging the eight distances determines $\zeta = 0.3142a$. The area $A = (3/2)a^2$ is the surface area of the cube of edge length $a/2$, which contains the volume of one f.u. In order to predict $a$ and $\zeta$ at higher pressures, theoretical pressure dependences of volume per f.u. ($V = a^3/8$) and the S-H distance $r_0 \equiv \zeta_1$ given in Ref. 14 are assumed and extrapolated to experimental values of pressure, as described in the Appendix. Values of $V$, $a$, and $\zeta$ calculated at experimental pressures $P$ are given in Table I, with uncertainties from extrapolating theoretical pressure dependences given in parentheses.

Assuming the maximum value $\sigma = 15/2 = 7.5$ for $CSH_7$ occurs at $P = 265 \pm 10$ GPa, which is in the range for maximum observed $T_C$, and with $a = 5.669 \pm 0.027$ Å and $\zeta = 1.755 \pm 0.009$ Å, one finds from Eq. (2) the result $T_{C0} = 283.0 \pm 3.5$ K, where the error bars combine the uncertainties of theoretical extrapolation and experimental pressure uncertainty. Given the uncertainties, $T_{C0}$ is in good agreement with experiment (Table I). Assuming the full complement of 6 electrons from sulfur would give 301.9 K. By averaging the two largest transition temperatures (questioned resistive datum omitted), to better account for the large uncertainties in pressure, a comparison between experiment and theory can be made showing remarkable equivalence with $\langle T_C \rangle = 282$ K and $\langle T_{C0} \rangle = 283$ K.

## B. Pressure dependence

Based on the abrupt superconductive resistance transitions observed for C-S-H over a broad range of applied pressures [1], bulk superconductivity is evidently maintained throughout. The transitions are quantified herein by defining the width $\Delta T_C$ as the difference between onset $T_C$ and the transition midpoint; results for $\Delta T_C$ so obtained are listed in Table I, along with the values of $P$, uncertainty $\Delta P$, and $T_C$, determined from Fig. 1(c) in Ref. 1. Deleted data from susceptibility are shown in gray font color with strikeouts. The questioned resistive datum is indicated by pink-colored font. Assuming that the $I\bar{4}3m$ $CSH_7$ structure is preserved, where $a$ and $\zeta$ vary smoothly with pressure, the precipitous decrease in $T_C$ with decreasing pressure below the maximum value must, therefore, be attributable to a decrease in the participating charge fraction and retained equilibrium between the two charge reservoirs. Consequently, the theoretical $T_{C0}$ function of Eq. (2) would be applicable to the experimental results for $T_C$ at various pressures, obtaining an experiment-based solution for $\sigma$ as,

$$\sigma^{(X)} = (3/2)[a\zeta T_C/\beta]^2, \quad (3)$$

where pressure-dependent $a$ and $\zeta$ are calculated from $I\bar{4}3m$ $CSH_7$ structure. Results for $\sigma^{(X)}$ calculated from the resistive $T_C$ according to Eq. (3) are plotted as a function of experimental pressure in Fig. 2, demonstrating that strong pressure dependence in $T_C$ is traceable to similar behavior in $\sigma^{(X)}$ (pink colored symbol denotes the questioned resistive datum). Owing to experimental uncertainties in measurements of pressure $\Delta P$, shown as horizontal error bars, uncertainties propagate into the calculations of $\sigma^{(X)}$. Such uncertainties are determined from the theoretical pressure dependences in $a$ and $\zeta$ as

$$\delta\sigma^{(X)} = (3/2)(T_C/\beta)^2 [\partial(a^2\zeta^2)/\partial P]\Delta P, \quad (4)$$

and are plotted as vertical error bars in Fig. 2.

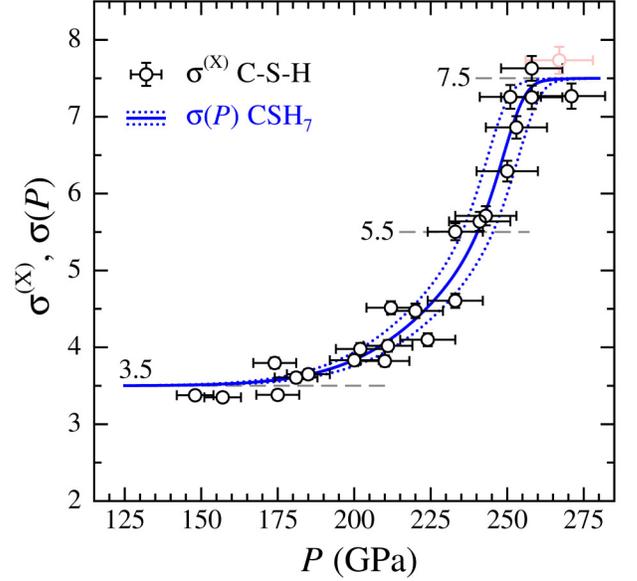

**FIG. 2.** The extracted $\sigma^{(X)}$ from resistance data (open circles) as defined in Eq. (3), plotted as a function of pressure $P$. The solid blue curve is a fit of the function $\sigma(P)$ from Eq. (5) to the data. The dotted curves represent the calculated uncertainty. The dashed horizontal lines at 3.5, 5.5, and 7.5 correspond to charge fractions for $H_3S$ (7/2), $H_3S + H_4$ (11/2), and $H_3S + CH_4$ (15/2), respectively. Excluded questioned resistive datum: pink color.

Theoretically, the participating charge fraction $\sigma$ equals one-half (from $\gamma = 1/2$) of the number of the participating valence electrons of the constituent atoms in the f.u. The likely origin of the unusually strong pressure dependence in $\sigma$ is the presence of $CH_4$, given that such behavior is not observed in $H_3S$ absent carbon [31]. The following two-component expression is introduced to model the pressure dependence,

$$\sigma(P) = (1/2)(n_S + 3n_H) + (1/2)\{4n_H(1 - \exp[-(P/P_1)^{q_1}]) + n_C(1 - \exp[-(P/P_2)^{q_2}])\}, \quad (5)$$

where the numbers of participating electrons are determined as $n_S = 4$ (sulfur), $n_H = 1$ (hydrogen), and $n_C = 4$ (carbon). The first term is the contribution from the 7 electrons in the $H_3S$ component and is presumed to be independent of pressure and the same as in $H_3S$ [27]. The second term models electrons contributed from $CH_4$, which are assumed to be non-participating at low pressure owing to localized electron density, e.g., as indicated for $P = 100$ GPa in Fig. S7(a) in the supplementary material in Ref. 24; electron transfer from $CH_4$ to $H_3S$ with increasing pressure is shown in Fig. S4 in the supplementary material of Ref. 24. Factors $[1 - \exp(-(P/P_i)^{q_i})]$ model the participation of $CH_4$-associated electrons, nearly zero at low pressure and approaching unity high pressures where the valence electrons become increasingly delocalized owing to reduced interatomic spacings. Such factors model the pressure dependence in analogy to the probabilities of electron tunneling through potential barriers that diminish dramatically at high pressure and are represented by power-law functions with large exponents. The parameters $P_1$, $P_2$, $q_1$, and $q_2$ are determined empirically from the resistive data (omitting the questioned datum) for $T_C$ vs. $P$ and theoretical pressure dependence of $a$ and $\zeta$. The continuous curve in Fig. 2 is the fit of Eq. (5) with parameters, $P_1 = 232(6)$ GPa, $P_2 = 249(3)$ GPa, $q_1 = 12(4)$, and $q_2 = 38(14)$. The dashed horizontal lines at 3.5, 5.5, and 7.5 correspond to charge fractions for $H_3S$ (7/2), $H_3S + H_4$ (11/2), and $H_3S + CH_4$ (15/2), respectively. The two dotted curves delimit uncertainties in $\sigma(P)$ originating from the experimental uncertainties in pressure and correspond to calculations of $\sigma(P \pm \Delta P)$.

According to Eq. (2), the fitted model function $\sigma(P)$ in Eq. (5) thus generates a prediction for the pressure dependence of $T_{C0}$ as,

$$T_{C0}(P) = [2\sigma(P)/3]^{1/2}\beta/a\zeta . \qquad (6)$$

Figure 3 shows $T_{C0}(P)$ as the solid blue curve, together with experimental resistive $T_C$ data (symbols) for $T_C$ vs. $P$ from Table I (experimental data from Ref. 1; questioned resistive datum in pink color). The two dotted curves denote uncertainties calculated from $T_{C0}(P \pm \Delta P)$. Taking the pressure uncertainties into account, the resistive data and theory differ by an average $(T_C - T_{C0}) = (-1.1 \pm 3.5)$ K; the root-mean-square difference between the symbols and solid curve in Fig. 3 is otherwise 5.8 K.

## III. DISCUSSION

The proposed structure of the C-S-H superconducting system is rather unique, comprising intercalated $CH_4$ molecules within an $H_3S$-like sublattice. What is even more interesting, from the perspective of the present work, is the sharp superconducting transition widths reported over a broad pressure range, indicative of a persisting charge balance between the two charge reservoirs throughout the pressure range. Taking the resistive data at face value, the analysis of $T_C(P)$ indicates that at the lowest pressure, only the $H_3S$ charges are involved in the superconductivity, with the number of participating charges increasing with pressure to include those of the $CH_4$ molecules. All three of these features make C-S-H a very special system.

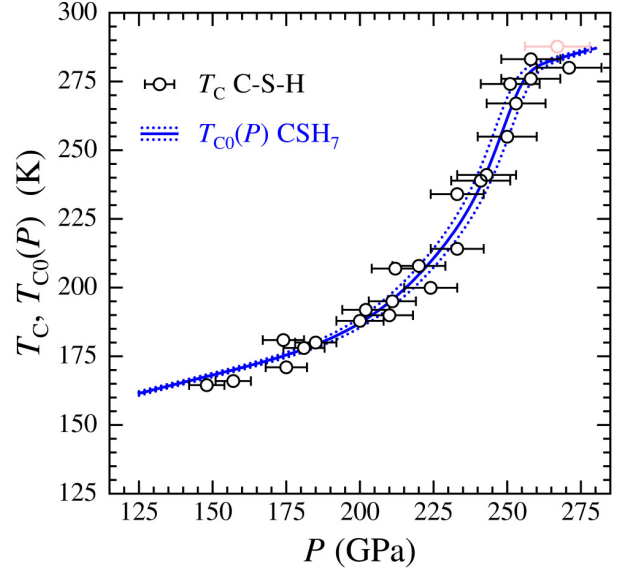

**FIG. 3.** Measured resistance transition temperature $T_C$ vs. pressure $P$ after Ref. 1 (open circles). The solid blue curve is $T_{C0}(P)$ from Eq. (6) to the data. The dotted blue curves represent the calculated uncertainty. Excluded questioned resistive datum: pink color.

Experimental data for C-S-H indicate limitations on the superconducting states. At the highest measured pressures shown in Fig. 3, the data are observed to approach a maximum $T_C$, suggesting an upper limit near 270 GPa (although more data are required for a definitive assessment). A lower bound in pressure is also indicated in Fig. 3, where data for $T_C$ (symbols) also fall systematically and progress-ively below the theoretical $T_{C0}$ (continuous curve) at pressures below about 175 GPa, which is particularly pronounced at the lowest reported pressure of 138 GPa. This may be due to a further reduction in participating charge (presumably from the $H_3S$ sublattice), or a consequence of differing crystal structure, given that orthorhombic symmetry is found by x-ray diffraction at pressures limited to below 178 GPa [19]. Superconducting phases are reported at lower pressures, depending on C content [32,33].

X-ray refinement of the *Pnma* structure at 143 GPa finds a sulfur arrangement with S-S distances of 2.401(7) to 3.004(6) Å [20], which may be compared to theoretical S-S distance $a/2 = 3.017$ Å, as determined from calculations of $I\bar{4}3m$ $CSH_7$ at this same pressure [24]. Thus, the framework of S atoms in the theoretical $I\bar{4}3m$ $CSH_7$ work bears close quantitative consistency with x-ray refinement data on C-S-H material, with caveats being the unresolved positions of light atoms H and C in an orthorhombic cell [20]. Theoretical compounds comprising intercalated $CH_4$ structures [23-26] also logically follow from the preferred syntheses reported for using methane as a precursor and the evident difficulties with elemental carbon [20]. Alternative theoretical structures containing $H_2$ tend to have lower theoretical phonon-mediated $T_C$, owing to the molecular characteristics of $H_2$ [24]; moreover, the reduced availability of participating electronic charges resulting from $H_2$ inclusions is also unfavorable for a wholly electronic mediation. By analogy in the case of $CSH_7$, the presence of $CH_4$ units in the form of localized molecules is similarly detrimental to not only phonon-mediated superconductivity [26], but also to the electronic mechanism should the electrons be non-participating.

The inability of electron-phonon theory to calculate adequately high values of $T_C$ for candidate compounds, which include C-doped $H_3S$ [20,21,23,24,34,35], suggests that phonon interactions are unlikely to be the primary pairing mechanism. Independently, evidence for unconventional superconductivity in C-S-H is also proposed through analysis of temperature dependence in normal-state resistance [36], similar to conclusions drawn for $H_3S$ and $LaH_{10}$ [37].

In identifying the pairing mechanism, a mass isotope effect may be considered within the context of observations from H-D isotopic substitutions in related $(H/D)_3S$ materials. Recent results report a mass isotope effect coefficient $\alpha \approx 0.22$ in $T_C$ from resistance measurements for $H_3S$ and $D_3S$ taken at nearly the same pressure, 155 and 157 GPa, respectively [38]. This result for $\alpha$ affirms the reanalysis [39] presented for earlier data [30,40], noting that the correction for the H-D isotopic shift in the pressure dependence of $T_C$ yields $\alpha = 0.13 \pm 0.11$ [39]. Given the added possibility of a sympathetic response of the lattice to an otherwise wholly electronic mechanism [27,30], the experimental $\alpha$ appears to fall well short of estimates for electron-phonon coupling [38]. Interpreting apparent shifts in $T_C$ also evidently entails taking into account mass isotope effects associated with the quantum nature of hydrogen bonding and structure [41]. Given the narrow superconductive transitions observed for C-S-H over a broad pressure range, and the absence of a phonon-based explanation of the elevated $T_C$, a companion study of C-S-D, assuming that high-$T_C$ C-S-D can be similarly synthesized, would greatly benefit the further understanding of these materials.

**TABLE I.** Experimental applied pressure $P$, uncertainty $\Delta P$, measured transition temperature $T_C$, and transition width $\Delta T_C$ are given for C-S-H from data in Ref. 1. Theoretical values (uncertainties) for volume per f.u. $V$, lattice parameter $a$, interaction and distance $\zeta$ are listed for $I\bar{4}3m$ $CSH_7$ at same values of $P$ (structural parameters from theory of Ref. 24). Gray strikeouts denote deleted susceptibility data. The questioned resistive datum [8,9] is shown in pink font.

| $P$ (GPa) | $\Delta P$ (GPa) | $T_C$ (K) | $\Delta T_C$ (K) | $V$ (Å$^3$) | $a$ (Å) | $\zeta$ (Å) |
|---|---|---|---|---|---|---|
| 271 | 11 | 280.0 | 9.6 | 22.60(14) | 5.655(11) | 1.734(4) |
| 267 | 10 | 287.7 | 0.5 | 22.74(14) | 5.666(11) | 1.738(4) |
| 258 | 10 | 276.0 | 0.4 | 23.04(14) | 5.691(11) | 1.746(4) |
| 258 | 10 | 283.1 | 1.2 | 23.04(14) | 5.691(11) | 1.746(4) |
| 253 | 10 | 267.0 | 2.1 | 23.22(14) | 5.706(11) | 1.751(4) |
| 251 | 10 | 274.1 | 5.5 | 23.29(14) | 5.711(11) | 1.752(4) |
| 250 | 10 | 254.9 | 2.1 | 23.32(14) | 5.714(11) | 1.754(4) |
| 243 | 10 | 241.0 | 0.3 | 23.58(14) | 5.735(11) | 1.761(4) |
| 241 | 10 | 238.9 |  | 23.65(14) | 5.741(11) | 1.763(4) |
| 233 | 9 | 234.0 | 7.8 | 23.95(14) | 5.765(11) | 1.771(4) |
| 233 | 9 | 214.1 |  | 23.95(14) | 5.765(11) | 1.771(4) |
| 224 | 9 | 199.9 |  | 24.29(14) | 5.792(11) | 1.781(4) |
| 220 | 9 | 207.9 | 1.3 | 24.45(14) | 5.805(11) | 1.785(4) |
| 212 | 8 | 206.9 |  | 24.77(14) | 5.830(11) | 1.794(4) |
| 211 | 8 | 195.0 |  | 24.81(14) | 5.833(11) | 1.795(4) |
| 210 | 8 | 189.9 | 0.6 | 24.85(14) | 5.836(11) | 1.796(4) |
| 202 | 8 | 191.9 |  | 25.18(14) | 5.862(11) | 1.806(4) |
| 200 | 8 | 187.9 |  | 25.27(14) | 5.869(11) | 1.808(4) |
| ~~189~~ | ~~8~~ | ~~197.9~~ |  | 25.75(14) | 5.906(11) | 1.821(4) |
| 185 | 7 | 180.0 |  | 25.93(15) | 5.920(11) | 1.827(4) |
| ~~182~~ | ~~7~~ | ~~191.1~~ |  | 26.07(15) | 5.930(11) | 1.830(4) |
| 181 | 7 | 178.0 |  | 26.11(15) | 5.933(11) | 1.832(4) |
| ~~178~~ | ~~7~~ | ~~184.4~~ |  | 26.25(15) | 5.944(11) | 1.836(4) |
| 175 | 7 | 171.0 |  | 26.40(15) | 5.955(11) | 1.840(4) |
| 174 | 7 | 180.9 | 0.2 | 26.45(15) | 5.959(11) | 1.841(4) |
| ~~166~~ | ~~7~~ | ~~172.0~~ |  | 26.84(15) | 5.988(11) | 1.852(4) |
| ~~160~~ | ~~6~~ | ~~169.9~~ |  | 27.14(15) | 6.011(11) | 1.860(4) |
| 157 | 6 | 166.0 |  | 27.30(15) | 6.022(11) | 1.865(4) |
| 148 | 6 | 164.5 |  | 27.79(15) | 6.058(11) | 1.878(4) |
| ~~138~~ | ~~6~~ | ~~146.8~~ |  | 28.36(15) | 6.099(11) | 1.894(4) |

The model genesis of the superconductivity in C-S-H comprises electronic interactions localized in real space, as depicted in Fig. 1, replicating earlier findings for $H_3S$, $LaH_{10}$ and $Cs_3C_{60}$ [27,29,30]. Calculations of $T_{C0}$ therefore apply to C-doped $H_3S$ materials [42], presuming that $\zeta$ and $A$ are similar to those calculated for the $CSH_7$ model and coupling among localized C-containing structures sufficiently sustains bulk superconductivity.

In Fig. 4, the average of the two highest values of observed $T_C$ for C-S-H (=282 K) is plotted against the average of $T_{C0}$ for $I\bar{4}3m$ $CSH_7$ (=283 K) calculated at the corresponding pressures, and compared to $T_C$ vs. $T_{C0}$ for several other optimally-compressed high-$T_C$ superconductors at the pressures indicated [27-30,43,44]. The solid line represents Eq. (1) and highlights the equality between measurements of $T_C$ and calculations of $T_{C0}$, with remarkable concurrence between theory and experiment exhibited over a range in $T_C$ covering nearly 300 K. A key takeaway from Fig. 4 is the common genesis of the superconductive state in both the new high-$T_C$ hydrides and the earlier cuprate-based and other high-$T_C$ materials. Since the universal constant $\Lambda$ in Eq. (1) approximates twice the reduced Compton wavelength ($\Lambda$ = 1.933(6)$\hbar/mc$), the involvement of Compton scattering in purely electronic high-$T_C$ pairing is considered plausible (see, e.g., Ref. 27).

The seminal work presented in Ref. 28 reports remarkable compliance with the methodology of 31 different superconductors from several superconducting families. Subsequent research increased this number, as reported in Ref. 27 in the case of $H_3S$, Ref. 29 in the case of the hydrogen clathrates, and Ref. 30 for $Cs_3C_{60}$. With the addition of $CSH_7$, this approach has been validated with statistical accuracy of ±1.26 K (or ±4.2%) in $T_{C0}$ for 56 different materials from 12 disparate superconducting families. These comprise the above-mentioned 3D compounds, layered cuprates, ruthenates, rutheno-cuprates, iron pnictides, BEDT-based [bis(ethylenedithio)tetrathia-fulvalene] organics, iron chalcogenides, intercalated group-4-metal nitride halides, and gated twisted bilayer graphene, with measured $T_{C0}$ values ranging from ~2 to 282 K.

## IV. CONCLUSION

Accepting an electronic genesis of the superconductivity in carbonaceous sulfur hydride (C-S-H), having a maximum resistive transition $T_C$ = 281–283 K at pressures $P$ = 258–271 GPa (Table I), a model is presented in which pairing is induced via Coulomb interactions between electronic charges occupying two sublattices, the H-S-H host network (type I, superconducting) and the $CH_4$ intercalate (type II, mediating). Assuming the computationally derived structure, $I\bar{4}3m$ $CSH_7$ [24], extrapolating $a$ and $\zeta$ to 265 GPa, and setting $\sigma$ = 7.5 (the maximum participating charge fraction), Eq. (2) yields $T_{C0}$ = 283.0 ± 3.5 K, which is in good agreement with data based on resistance drops [45]. Averaging the two highest transition temperatures, a remarkable equivalence between experiment and theory is found with $\langle T_C \rangle$ = 282 K and $\langle T_{C0} \rangle$ = 283 K.

Most curious, however, is the anomalously narrow transition widths $\Delta T_C$ observed for C-S-H over a wide pressure range, indicating that charge equilibration between the two charge reservoirs is maintained throughout. This revelation, coupled with the behavior of $T_C(P)$ shown in Fig. 1(c) of Ref. 1, necessitates a pressure-dependent $\sigma$ ranging between 3.5 and 7.5, with essentially all of the variation in $\sigma$ originating from $CH_4$ electrons via pressure-induced dispersion of the C–H bonds. Functional forms for $\sigma(P)$ and $T_{C0}(P)$, assuming tunneling behavior of the $CH_4$ electrons, are derived and shown to compare well with resistive data (omitting questioned datum) for $T_C(P)$. Combined with experimental uncertainties in pressure, variations in $\sigma$ yield overall differences averaging ($T_C$ − $T_{C0}$) = (−1.1 ± 3.5) K.

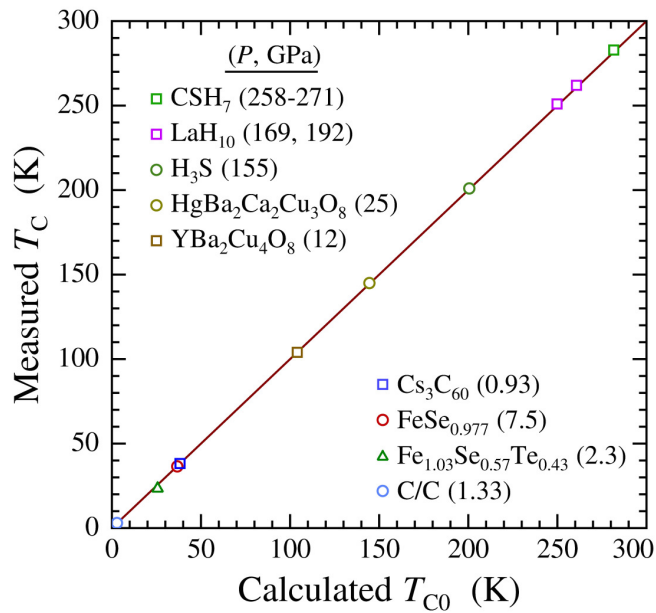

**FIG. 4.** Measured $T_C$ vs. calculated $T_{C0}$ for compressed high-$T_C$ superconductors; $I\bar{4}3m$ $CSH_7$ (258-271 GPa), $Fm\bar{3}m$ $LaH_{10}$ (169 and 192 GPa) [29], $Im\bar{3}m$ $H_3S$ (155 GPa) [27], $HgBa_2Ca_2Cu_3O_{8+\delta}$ (25 GPa) [28], $YBa_2Cu_3O_8$ (12 GPa) [29], A15 $Cs_3C_{60}$ (0.93 GPa) [30], $FeSe_{0.977}$ (7.5 GPa) [43], $Fe_{1.03}Se_{0.57}Te_{0.43}$ (2.3 GPa) [43], and C/C (twisted bilayer graphene device D2, 1.33 GPa) [44]. The solid line represents Eq. (1), highlighting the equality between $T_C$ and $T_{C0}$.

## ACKNOWLEDGMENTS

The authors are grateful for support from the College of William and Mary, New Jersey Institute of Technology, and the University of Notre Dame.

## AUTHOR DECLARATIONS

### Conflict of Interest

The authors declare that they have no conflict of interest.

## DATA AVAILABILITY

The data that support the findings of this study are available within the article.

## APPENDIX: PRESSURE DEPENDENCE OF STRUCTURAL PARAMETERS

Calculations for $I\bar{4}3m$ $CSH_7$ at pressures up to 200 GPa are presented in Ref. 24, yielding pressure dependences of volume per $H_3S$ host lattice ($V$) in Fig. S9 and various S-H spacings in Fig. S8 in the supplementary material. Pressure dependences in $a$ and $\zeta$ are determined by modeling these results.

Transcribed calculations of $V$ vs. pressure $P$ are shown as symbols in Fig. 5(a). The pressure dependence is fitted to a model emulation of the equation of state, $V(P) = [c_0 + c_1 P + c_2 \exp(-c_3 P)]^{-1}$, with parameters $c_0 = 0.0268(2)$ Å$^{-3}$, $c_1 = 6.44(8) \times 10^{-5}$ Å$^{-3}$GPa$^{-1}$, $c_2 = -0.0080(1)$ Å$^{-3}$, and $c_3 = 0.0210(7)$ GPa$^{-1}$ with root-mean-square fitting error of 0.016 Å$^3$, and is shown by the solid curve in Fig. 5(a). Evaluations of $V(P)$ and lattice parameters $a = (8V)^{1/3}$, along with the fitting uncertainties, are given for the various experimental pressures in Table I.

Transcribed calculations for the S-H distance $r_0$ (illustrated as $\zeta_1$ in Fig. 1), divided by $a$ and normalized to $r_0/a$ at pressure $P_{100} \equiv 100$ GPa, are shown as symbols in Fig. 5(b). The red curve is the fitted model function $[r_0/a](P) = b_1 + b_2 \exp(-b_3 P)$ with parameters $b_1 = 0.9745(8)$, $b_2 = 0.0720(6)$, and $b_3 = 0.0104(3)$ GPa$^{-1}$ with root mean square fitting error of $3.1 \times 10^{-4}$. Pressure dependence in $\zeta$ is scaled to calculations at $P = P_{100}$ as $\zeta(P) = (\zeta/a)|_{P_{100}}\, a\, [r_0/a](P)$, noting that $r_0/a$ decreases by only 2.1 % from 100 to 271 GPa. Pressure dependent $\zeta$ values are given in Table I.

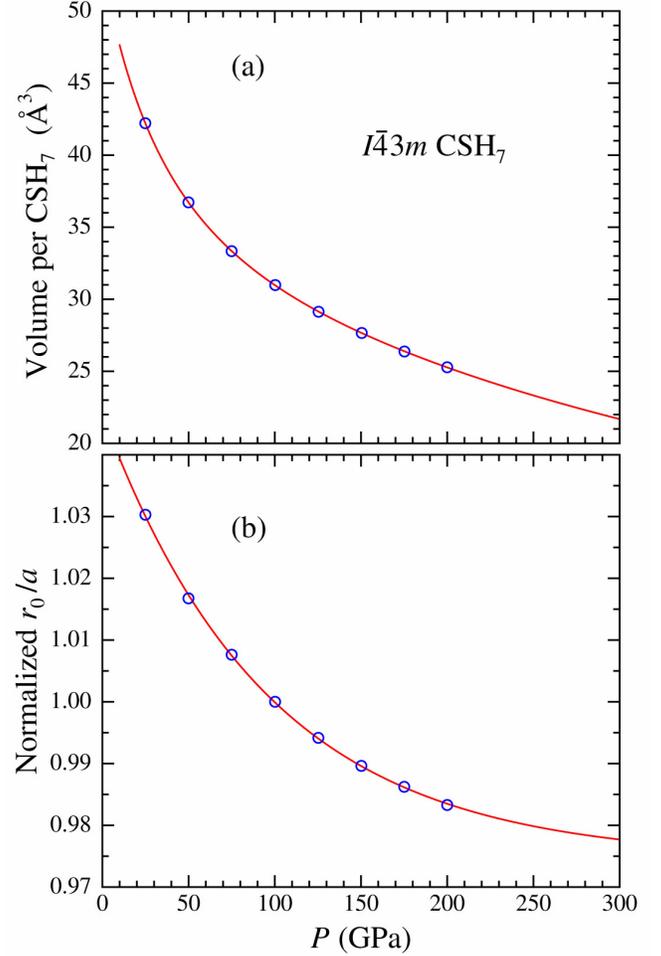

**FIG. 5.** Pressure dependence of theoretical (a) Volume per f.u. and (b) $r_0/a$ (normalized to its value at $P = 100$ GPa) for $I\bar{4}3m$ $CSH_7$ after Figs. S9 and S8 in the supplementary material of Ref. 24, plotted as a function of pressure $P$. The curves are model functions described in the text.